\documentclass[10pt,twocolumn,superscriptaddress,aps,pra,showpacs,preprintnumbers,amsmath,amssymb,floatfix]{revtex4}

\usepackage[latin9]{inputenc}
\usepackage{color}

\usepackage{braket}
\usepackage{graphicx}
\usepackage{esint}
\usepackage[unicode=true,  bookmarks=false,  breaklinks=false,pdfborder={0 0 1},backref=false,colorlinks=true]  {hyperref}

\makeatletter
\@ifundefined{textcolor}{}
{%
 \definecolor{BLACK}{gray}{0}
 \definecolor{WHITE}{gray}{1}
 \definecolor{RED}{rgb}{1,0,0}
 \definecolor{GREEN}{rgb}{0,1,0}
 \definecolor{BLUE}{rgb}{0,0,1}
 \definecolor{CYAN}{cmyk}{1,0,0,0}
 \definecolor{MAGENTA}{cmyk}{0,1,0,0}
 \definecolor{YELLOW}{cmyk}{0,0,1,0}
}

\usepackage{graphicx}
\usepackage[english]{babel}
\usepackage{amsmath}
\usepackage{amssymb}
\definecolor{orange}{rgb}{1,0.5,0}
\usepackage{bm}

\newcommand{\bx}{{\mathbf x}}

\newcommand{\bk}{{\mathbf k}}
\newcommand{\sx}{\sigma^x}
\newcommand{\sy}{\sigma^y}
\newcommand{\sz}{\sigma^z}
\newcommand{\sd}{\sigma^-}
\newcommand{\su}{\sigma^+}

\newcommand{\tr}{\mbox{Tr}}
\newcommand{\z}{\mathfrak z}

\begin{document}
\title{Nonequilibrium many-body steady states via Keldysh formalism}
\author{Mohammad F. Maghrebi}
\email[Corresponding author: ]{magrebi@umd.edu}
\affiliation{Joint Quantum Institute, NIST/University of Maryland, College Park, Maryland 20742, USA}
\affiliation{Joint Center for Quantum Information and Computer Science, NIST/University of Maryland, College Park, Maryland 20742, USA}
\author{Alexey V. Gorshkov}
\affiliation{Joint Quantum Institute, NIST/University of Maryland, College Park, Maryland 20742, USA}
\affiliation{Joint Center for Quantum Information and Computer Science, NIST/University of Maryland, College Park, Maryland 20742, USA}

\begin{abstract}
Many-body systems with both coherent dynamics and dissipation constitute a rich  class of models which are nevertheless much less explored than their dissipationless counterparts.
The advent of numerous experimental platforms that simulate such dynamics poses an immediate challenge to systematically understand and classify these models. In particular, nontrivial many-body states emerge as steady states under non-equilibrium dynamics. While these states and their phase transitions have been studied extensively with mean field theory, the validity of the mean field approximation has not been systematically investigated.
In this paper, we employ a field-theoretic approach based on the Keldysh formalism to study nonequilibrium phases and phase transitions in a variety of models. In all cases, a complete description via the Keldysh formalism indicates a partial or complete failure of the mean field analysis. Furthermore, we find that an effective temperature emerges as a result of dissipation, and the universal behavior including the dynamics near the steady state is generically described by a thermodynamic universality class.

\end{abstract}

\maketitle

\begin{table*}
\begin{tabular}{|c||c|c|c|} \hline
   & \begin{tabular}{c}
   \\
     Model
     \\
     \\
   \end{tabular}  & Mean Field & Field Theory\\
\hline \hline
   \quad A \qquad  & \begin{tabular}{c}
   \\
   \(
    H=-J \sum_{\langle i j\rangle }\sx_i \sx_j+\Delta \sum_i\sz_i
   \)
   \\
   \\
   \(
   L_i=\sqrt{\Gamma}\,\sd_i
   \)
   \\
   \\
   \end{tabular}
   &
   \begin{tabular}{c}
   Continuous transition from
   \\
   an ordered phase ($\langle\sx_i\rangle = $const) to
   \\
   a disordered phase ($\langle\sx_i\rangle=0$)
   \end{tabular}
   &
   \begin{tabular}{c}
   First order transition
   \\
   at sufficiently strong dissipation
   \end{tabular}
   \\
   \hline \hline
   \quad B \qquad  & \begin{tabular}{c}
   \\
   \(
H=-J \sum_{\langle i j\rangle }\sigma^+_i \sigma^-_j
\,\,+\Omega \sum_i\sigma^x_i+\Delta \sum_i \sigma^z_i
   \)
   \\
   \\
   \(
   L_i=\sqrt{\Gamma}\,\sd_i
   \)
   \\
   \\
   \end{tabular}
   &
   \begin{tabular}{c}
   Continuous transition from
   \\
   a phase with one stable solution to
   \\
     a bistable regime
   \end{tabular}
   &
   \begin{tabular}{c}
      No bistability
   \end{tabular}
   \\ \hline \hline
   \quad C \qquad  & \begin{tabular}{c}
   \\
   \(
H=J \sum_{\langle i j\rangle } \,\, \sx_i \sx_j- \sy_i \sy_j
   \)
   \\
   \\
   \(
   L_i=\sqrt{\Gamma}\,\sd_i
   \)
   \\
   \\
   \end{tabular}
   &
   \begin{tabular}{c}
   Continuous transition from
   \\
   an XY phase (staggered $\langle\su_i\rangle$) to
   \\
   a disordered phase ($\langle\su_i\rangle=0$)
   \end{tabular}
   &
   \begin{tabular}{c}
   No XY phase in
   \\
   $d=2$ dimensions
   \end{tabular}
   \\\hline \hline
   \quad D \qquad  & \begin{tabular}{c}
   \\
   \(
    H=-J \sum_{\langle i j\rangle }\sigma^+_i \sigma^-_j \,\,+ \Delta \sum_i \sigma^z_i
   \)
   \\
   \\
   $L^l_i=\sqrt{\Gamma}\,\sd_i$ \qquad
   $L^p_i=\sqrt{\Gamma_p}\,\su_i$ \\
   \null \hskip -40pt $L^I_{ij}=\sqrt{\kappa} \, (1+\sz_j)\,\sd_i$
   \\
   \\
   \end{tabular}
   &
   \begin{tabular}{c}
   No phase transition
   \\
   or spontaneous symmetry breaking
   \\
   ($\langle\sx_i\rangle=\langle\sy_i\rangle=0$)
   \end{tabular}
   &
   \begin{tabular}{c}
   A continuous transition to a
   \\
   spontaneously symmetry broken
   \\
   phase ($\langle\su_i\rangle=$ const)
   \end{tabular}
   \\
   \hline
\end{tabular}
\caption{Summary of the results. We consider four driven-dissipative models with spin-1/2s on a $d$-dimensional cubic lattice. In each case, the Hamiltonian ($H$) and the dissipation via the Lindblad operators ($L$) are defined. The mean-field theory prediction is described and contrasted with the Keldysh field-theoretic treatment. Nearest-neighbor Ising, isotropic XY, and anisotropic XY interactions are considered in different examples. In models (A), (B), and (C), we take $d=2$ or $3$, while, in model (D), we take $d=3$. In the former three models, the dissipation is an independent spontaneous emission on each site, whereas, in the latter model, pump and correlated dissipation on nearby sites are assumed as well. The interplay of unitary and dissipative dynamics creates rich phase diagrams. Our analysis indicates that the mean-field predictions are incomplete or even wrong for all four models, and should be supplemented with field theory via the Keldysh formalism. \label{tab1}}
\end{table*}

\section{Introduction}

Condensed matter systems typically relax to their equilibrium state on very short time scales. A successful paradigm of statistical physics developed over many years explains various aspects of equilibrium or near-equilibrium phenomena in many-body systems. Specifically, quantum phase transitions, which emerge at zero temperature, have been the subject of intense research over the past several decades \cite{SachdevBook}.

In contrast, experiments with ultracold matter have opened new avenues to probe far-from-equilibrium many-body systems in the presence of both coherent dynamics and controlled dissipation, the so-called driven-dissipative systems. While the subject is not new, the vast range of experimental platforms have brought driven-dissipative many-body systems into light again
{and made it necessary to undertake} a more thorough investigation.
Experiments range from polariton condensates in the context of semiconductor quantum wells
in optical cavities \cite{Kasprzak06,Lagoudakis08,Roumpos12,Moskalenko2000,Ciuti13-2} and arrays of microcavities \cite{Clarke08,Hartmann08,Houck12,Schmidt13} to trapped ions \cite{Blatt12,Britton12} and optomechanical  setups \cite{Marquardt09,Chang11,Ludwig13}. Furthermore,
experiments on strongly interacting Rydberg polaritons are already probing non-equilibrium many-body physics \cite{Pritchard10,Dudin12,Peyronel12,gunter13},
while cavity-quantum-electrodynamics experiments can potentially explore dynamics under exotic many-body Hamiltonians with glassy ground states
\cite{Strack11,Gopalakrishnan11,Muller12}.
The interplay between dissipation, which is generically present in these systems, and coherent dynamics leads to new, and inherently nonequilibrium, phases.
The fundamental question is then how we should understand and classify these phases. Do dissipative-driven systems give rise to new phases of matter? What are the universal properties of the associated phase transitions?

The sophisticated toolbox of equilibrium physics is not immediately applicable to nonequilibrium problems. The equivalents of equilibrium concepts such as temperature, free energy, and partition function either have no obvious counterpart out of equilibrium, or often become intractably complicated. In particular, the main goal of condensed matter physics is to find properties of the ground state or, at finite temperature, the thermal state of the system, while, out of equilibrium, nontrivial many-body states emerge as steady states under non-equilibrium dynamics.
In the absence of
{a powerful}
systematic approach, approximations such as mean field theory have been widely used \cite{Littlewood07,Diehl08,Diehl10,Diehl11, Lee11,Lee12,Lee13,Ciuti13,Marcuzzi14}, but are often at odds with other analytical and numerical studies \cite{Weimer15,Weimer15-2}, sometimes even in the limit of infinite dimensions \cite{Pohl14} where the equilibrium mean field is known to be exact;
these analytical and numerical studies are based on variational methods~\cite{Weimer15,Weimer15-2} and approximate rate equations \cite{Pohl14}.
{In general, numerical techniques are either limited to one dimension, e.g.\ t-DMRG \cite{Pichler10}, or to infinite dimensions such as nonequilibrium dynamical mean field theory \cite{Freericks06,Aoki14}, or cannot be applied to non-equilibrium systems, e.g.\ path-integral
Monte Carlo \cite{ceperley95}.}
It is worth mentioning that there exist exact solutions, due to integrability, for a very specific class of nonequilibrium models where the system is driven only at the boundaries \cite{Prosen14,Prosen15}.

In principle, the Keldysh-Schwinger functional integral formalism provides {a} general approach to nonequilibrium physics. A notable example is the universal behavior of early evolution of an initial state prepared far from equilibrium \cite{Janssen89}, see also \cite{Calabrese05} for a review. This approach has been applied to a number of driven-dissipative systems such as lossy polariton condensates \cite{Sieberer13,Sieberer14,Altman15} and driven atomic ensembles interacting with a cavity mode \cite{DallaTorre13}. Indeed, a systematic application of the Keldysh formalism to the wide variety of driven-dissipative problems is far from complete. Surprisingly, even the simplest many-body driven-dissipative spin models have not been fully tackled with the Keldysh formalism; these models should not be confused with spin-boson models where a two-level system is coupled to a dissipative environment in thermal equilibrium \cite{Leggett87}. To be more precise, in contrast to spin-boson models where spins are \emph{strongly} coupled to a thermal environment, driven-dissipative systems studied in this paper deal with situations where coupling to the environment is \emph{weak}, but the system is driven out of thermal equilibrium by an external drive (for example, by a laser beam), or it may be in contact with two different baths at different temperatures, or may be coupled to a non-thermal bath.

In this paper, we consider a variety of nonequilibrium models, mostly spin models on a $d$-dimensional cubic lattice, and study their, inherently nonequilibrium, steady states.
The fact that we are interested in the near-critical
{(criticality identified by a diverging time scale)} long-distance behavior of the models allows us to map the spin models to continuum field theories, which we study via the Keldysh formalism in great detail. For each model, we compare and contrast the field-theoretic Keldysh approach to the mean field theory which is shown to miss some or most of the features of the full field-theory treatment.
A unified and systematic Keldysh approach is applied to
{most of} these models
{(Models A-C in Table \ref{tab1})}: Close to the mean-field phase transition, critical and massive components of the field are identified, and the latter is integrated out to find an effective action for the critical field.
{While such a procedure for obtaining an effective action is}
standard,
{the Keldysh formalism}
involves two different, classical and quantum, fields which require special care.

We first provide a brief description of the tools used throughout this paper in Sec.~\ref{Sec. Framework}, and then study four specific driven-dissipative models in Sec.~\ref{Sec. DD Models}. We generically find that a driven-dissipative model behaves thermally, that is, an effective temperature emerges, and the phase transition between different phases is described by well-known thermodynamic paradigms and their universality classes. The emergence of an effective equilibrium and a conventional thermal phase transition has been identified in many contexts \cite{mitra06,wouters06,dalla-torre10,Diehl10,Gopalakrishnan11,oztop12,Sieberer13,DallaTorre13,Sieberer14,Diehl14,Sieberer15}. Furthermore, effects beyond mean field have been studied in the context of driven-dissipative condensates \cite{Sieberer13,Sieberer14,Altman15}.
Nevertheless, more generally, the determination of what  precise thermodynamic phase corresponds to the nonequilibrium steady state is often nontrivial and depends sensitively on the type of dissipation and its competition with the coherent dynamics. Systematic derivation of this correspondence for each of the models under consideration constitutes the main result of the present manuscript. What is universal in all these models is the emergence of thermal phase transitions, but they show generic, although model-dependent, effects of fluctuations in driving phase transitions to a different order (model A), removing mean-field artifacts (models B, C), melting order (model C), or inducing symmetry breaking (model D).

For the benefit of the reader,
we summarize our results
in Table \ref{tab1}, which
introduces the four driven-dissipative models under consideration and highlights the important differences between the results obtained via mean field theory and via the Keldysh field-theoretic formalism.

\section{General framework}\label{Sec. Framework}
In this section, we closely follow Refs.\ \cite{Littlewood07,DallaTorre13,Sieberer13,Sieberer14} to introduce the general framework in which we define and study driven-dissipative systems; see also the pioneering works in Refs.~\cite{Martin73,DeDominicis87,Janssen76}. To properly treat dissipation, we first introduce  a second-quantized master equation under which the density matrix evolves with both unitary and dissipative dynamics. We briefly sketch how the master equation is mapped to the Keldysh path integral. We then consider a generic form of the Keldysh action, and present a simple scaling argument that constitutes the first step of the RG procedure. Finally, we show that, under certain conditions, the Keldysh path integral maps to a classical Langevin equation, and the resulting steady state can be expressed as a classical partition function. While the tools and methods discussed in this section are known and have been utilized in the literature, we find it useful to present them in an introductory section rather than an appendix as we will use them frequently, and mostly in the context of models to which they have not been directly applied.
{For a detailed introduction to the Keldsyh formalism, we refer the reader to Refs.~\cite{KamenevBook,Kamenev09}. The application of the Keldysh formalism to a number of driven-dissipative systems can be found in Refs.~\cite{DallaTorre13,Sieberer13,Sieberer14}, which this work has greatly benefitted from.
}

  \emph{Quantum master equation.}---To describe an open system, one should include both the coherent and dissipative processes in a master equation for the density matrix as (in units where $\hbar=1$)
  \begin{equation}\label{Eq. Master equation}
    \partial_t \rho =-i [H, \rho] +\frac{1}{2} \sum_\alpha \left(2 L_\alpha \rho L_\alpha^\dagger - L_\alpha^\dagger L_\alpha \rho- \rho L_\alpha^\dagger L_\alpha \right).
  \end{equation}
   The first term on the right-hand side gives the usual coherent evolution via the Hamiltonian $H$.
   The dissipation is subsumed in the second term (sometimes referred to as the linear Liouville operator ${\cal L}[\cdot]$ acting on $\rho$) characterized by the so-called Lindblad operators $L_\alpha$ which describe an incoherent process $\alpha$.
   The master equation relies on the Born-Markov approximation, which assumes that the reservoir has a short correlation time and is large enough to be unaffected by the coupling to the system
   \cite{Gardiner2004}.
  In the so-called stochastic-wavefunction interpretation, the operator $L_\alpha$ acts as an occasional,  discontinuous, jump from one state to another as a result of an incoherent process \cite{dalibard92} (more precisely, the first term in parentheses in Eq.\ (\ref{Eq. Master equation}) describes such a jump, while the last two terms take dissipation into account between the jumps).
  Typical Lindblad operators are local operators causing transitions between levels or decay of coherences, i.e.\ dephasing.
   Various examples of Hamiltonian dynamics and Lindblad operators are discussed in Sec.~\ref{Sec. DD Models} and summarized in Table \ref{tab1}.

   \emph{Keldysh functional integral.}---The Keldysh formalism provides a general framework to study nonequilibrium problems with functional integral techniques. Within this approach, the action is defined on a closed time contour with a forward and a backward branch. In performing the functional integration, one should sum over all configurations with independent values of the underlying fields on the two branches. To be concrete, consider $\{a\}$ as a shorthand for all the fields in a particular model. The Keldysh path integral can be formally expressed as
   \begin{equation}\label{Eq. Keldysh fn int}
     \int {\prod}_{\sigma=\pm} \!\!D\left[a_\sigma(t,\bx)\right] \,\, e^{i S_K[a_+,a_-]},
   \end{equation}
   where $\sigma=\pm$ denotes the forward and backward branches, respectively, while $(t,\bx)$ are the time and spatial coordinates.
   Evaluating the functional integral on a closed time contour means that the values of $a_{\pm}$ should match at $t=\infty$
   (at $t=-\infty$, the system is described by an initial state whose precise form is unimportant for the steady state of the system at long times \cite{KamenevBook}, although the early evolution of the system depends sensitively, and perhaps even universally, on the initial state \cite{Janssen89}). All the information in the, possibly time-dependent, state and specifically all correlation functions can be computed by inserting fields in the functional integration; for a general operator $\hat O$,
   one has
   \begin{equation}
     \tr\left[\rho(t) \hat O \right] =\left \langle O_+(t) \right\rangle\,,
   \end{equation}
   where the expectation value $\langle \cdot\rangle$ is computed with respect to the functional integral in Eq.~(\ref{Eq. Keldysh fn int}). The subscript $+$ on $O$ on the right-hand side of the above equation implies that all the fields inside $O$ are evaluated on the forward branch; one can equally well arrange to have the observable on the backward branch, but it is, in fact, often more convenient to work in a different basis, which we shall define shortly \cite{KamenevBook,Kamenev09}.

   Conveniently, the master equation (\ref{Eq. Master equation}) can be directly mapped to a Keldysh action comprising both the coherent ($H$) and dissipative ($D$) terms as
   \begin{equation}
       S_K = S_H + S_D\,.
   \end{equation}
   The coherent part of the action can be written in the coherent-state representation
   ({after normal ordering, and} assuming that $a$ corresponds to a bosonic operator) of the path integral as
   \begin{equation}\label{Eq. SH}
    S_H= \sum_{\sigma=+,-} \sigma \left[\left(\int_{t,\bx} a^*_\sigma i \partial_t a_\sigma \right) - H[a_\sigma, a^*_\sigma] \right]\,.
   \end{equation}
   The relative sign of the forward and backward branches has its origin in the minus sign in the commutator $[H,\rho]$.
   The dissipative dynamics yields the Keldysh term
   \begin{equation}\label{Eq. SD}
     S_D=-i \sum_\alpha \int_{t,\bx} L_{\alpha,+}L_{\alpha,-}^* -\frac{1}{2} \left( L_{\alpha,+}^*L_{\alpha,+}+L_{\alpha,-}^*L_{\alpha,-}\right)\,,
   \end{equation}
   with the Lindblad terms $L_\alpha$ and $L_\alpha^*$ given in terms of position- and time-dependent fields $a_\sigma$ (see Ref.~\cite{Sieberer14} for more details). In general, operators acting on the density matrix $\rho$ from the left (right) give rise to a corresponding term on the $\sigma=+$ ($\sigma=-$) contour \cite{DallaTorre13,Sieberer14}.

   It is often more convenient to work in the Keldysh basis defined as
   \begin{equation}\label{Eq. Keldsyh basis}
    a_{cl}= \frac{a_++a_-}{\sqrt{2}}\,, \quad a_{q}= \frac{a_+-a_-}{\sqrt{2}}\,,
   \end{equation}
   where $a_{cl/q}$ are the so-called classical and quantum fields; typically, in an ordered phase $\langle a_{cl}\rangle =$ const, while $\langle a_q\rangle=0$, that is, the quantum field is purely fluctuating.

  \emph{Scaling dimensions.}---We often encounter the Keldysh action of the form $\int_{t,\bx}  a_q^* (i\partial_t+ \nabla^2-i\frac{\Gamma'}{2})a_{cl}+ i\Gamma |a_q|^2$ at the quadratic level, where $\Gamma$ and $\Gamma'$ generally designate dissipation rates in the model. To find the scaling dimension of the fields as a first step of the RG procedure, we perform a simple dimensional analysis; setting the scaling dimension of spatial and time derivatives as  $[\nabla]=1$ and $[\partial_t]=z$, we have $z=2$ at the quadratic order. We also choose $\mbox{dim}[\Gamma]=0$. The scaling dimensions of the classical and quantum fields are then \cite{Sieberer13,Sieberer14}
  \begin{equation}
     [a_{cl}]=\frac{d-2}{2}\,, \qquad
     [a_q]=\frac{d+2}{2}\,,
  \end{equation}
   with $d$ the number of spatial dimensions. Notice that, with our choice of renormalization, the term $\int_{t, \bx}|a_q|^2$ is only marginally relevant; any additional powers of $a_{cl/q}$ or higher-order derivatives in the integrand make it irrelevant \footnote{A simple dimensional analysis gives $[a_{cl}]=0$ in $d=2$ dimensions; however, the  classical field acquires a nonzero scaling dimension in the course of the RG.}. Therefore, the quantum field $a_q$ appears at most quadratically, and without any derivatives at the quadratic order, in the relevant part of the action.
   {We stress that our analysis based on power counting is valid at or near criticality (criticality defined as $\Gamma'=0$).}

   \emph{Langevin equation.}---{In the models presented in this paper, we often find a Keldysh path integral of the form
   \begin{align}\label{Eq. Keldysh fn q-quadratic}
     & \int  D\!\left[a_{cl/q}(t,\bx)\right] e^{iS_K[a_{cl},q_q]}\quad \mbox{with}\nonumber \\
      &S_K=\int_{t,\bx} a_q^*\left[i\, \dot  a_{cl}+\frac{\delta f[a_{cl}]}{\delta a_{cl}^*}\right]+c.c. +i \Gamma |a_q|^2\,,
   \end{align}
   where the quantum field appears at most at the quadratic level, and}
   $f$ is assumed to be a complex-valued functional of $a_{cl}$;
   { $f$ may generically include gradients of the field as well as interaction terms}.
   {By the virtue of a Hubbard-Stratonovich transformation, one  can cast the quadratic term in the quantum field in terms of a `noise' field $\xi(t,\bx)$. The fluctuations of the classical field in the above path integral can be then cast \emph{exactly} in terms of a classical Langevin equation as} \cite{KamenevBook}
   \begin{equation}\label{Eq. Langevin}
     i\partial_t a_{cl}(t,\bx)= -\frac{\delta f[a_{cl}]}{\delta a_{cl}^*}+\xi(t,\bx),
   \end{equation}
   where noise correlations satisfy
   \begin{equation}
     \langle \xi(t,\bx) \xi^*(t',\bx')\rangle  = \Gamma\, \delta(\bx-\bx') \delta(t-t')\,.
   \end{equation}
   {All correlations of the classical field can be computed either via the Keldysh path integral, Eq.~(\ref{Eq. Keldysh fn q-quadratic}), or equivalently via the Langevin equation (\ref{Eq. Langevin}). The latter, however, has the obvious advantage of mapping to a classical stochastic equation. We are often interested in finding the steady state of
   Eq.~(\ref{Eq. Langevin}), which,
   via a series of steps outlined above, is indeed equivalent (for the purposes of probing the long-wavelength physics near criticality) to the steady state of the quantum master equation~(\ref{Eq. Master equation}). The calculation of the steady state of Eq.~(\ref{Eq. Langevin})  is not trivial in general. However, }for the {
   special} case{---encountered
   {in Models A-C in Table \ref{tab1}}---}
   where the function $f$ is purely imaginary \footnote{For a more general case, see Refs.~\cite{Deker75,Graham90}.}, $f[a_{cl}]=if_I[a_{cl}]$, the steady state is given by the probability distribution
   \begin{equation}\label{Eq. PF}
     P[a_{cl}] \sim \exp\left(-\frac{f_I[a_{cl}]}{T_{\rm eff}}\right),
   \end{equation}
   which takes the form of a thermodynamic partition function with the effective temperature $T_{\rm eff}\equiv\Gamma/2$, and $f_I$ as the effective classical Hamiltonian of the system.
   The value of the effective temperature is chosen such that the fluctuation-dissipation condition, relating the fluctuations of the noise to the dissipation in the effective model, holds.
   {Computing correlation functions weighted by Eq.~(\ref{Eq. PF}) will produce all the relevant information in the quantum steady state.}
   We also remark that Eq.~(\ref{Eq. PF}) can be derived from the \emph{Fokker-Planck }equation that casts the Langevin equation (\ref{Eq. Langevin}) in the form of an equation for the probability distribution  \cite{KardarBook}.

   {We stress that driven-dissipative systems may not always be described by a thermal-like probability distribution. In particular, this description may fail if the system is not invariant under a symmetry transformation that characterizes the equilibrium condition \cite{Sieberer14,Sieberer15}.
    }

    {Although the procedure and the steps outlined in this section are rather standard, the key challenge lies in bringing a many-body model of interest, e.g.\ a spin model, into the form given by Eq.~(\ref{Eq. Keldysh fn q-quadratic}) or Eq.~(\ref{Eq. Langevin}); in almost all the models studied in this paper (Models A-C in Table \ref{tab1}; Model D requires a special treatment), we find that such a transformation is nontrivial. Nevertheless, we show how to achieve this goal via a systematic approach.}

\section{Driven-dissipative models}\label{Sec. DD Models}
Here, we consider a number of driven-dissipative models, mostly spin systems on a cubic lattice, each defined by a particular Hamiltonian and a particular form of dissipation. Several models considered here are novel and have not been studied in the literature even at the mean-field level.
In each subsection, we compare the results of the mean-field analysis to those of the full field-theoretic treatment using the Keldysh formalism. The mean-field analysis is shown to partially or completely fail in all the models, while the Keldysh approach provides a field-theoretic formalism to go beyond mean field,
 best suited to study the vicinity of phase transitions.

The driven-dissipative spin models described below can be implemented using a variety of experimental systems. The Hamiltonians can be implemented using ions coupled via motional modes \cite{molmer99,porras04,kim10,richerme14,jurcevic14}, atoms in optical lattices coupled via superexchange \cite{duan03,pinheiro13,fukuhara13}, atoms in optical cavities or along waveguides coupled via optical modes \cite{Gopalakrishnan11,douglas15,gonzalez-tudela15}, or  polar molecules \cite{barnett06,micheli06,gorshkov11b,gorshkov11c,gorshkov13,yan13}, Rydberg atoms \cite{barredo15}, magnetic atoms \cite{paz13}, magnetic defects in solids \cite{dolde13}, and Rydberg polaritons \cite{Firstenberg13} coupled via dipole-dipole or van der Waals interactions.
The Lindblad operators either arise in these models naturally via processes like spontaneous emission or can be engineered via optical pumping.

  \subsection{Transverse-field Ising model (TFIM) with spontaneous emission in the eigenbasis of the field}\label{Sec. TFIM}
  In this subsection, we consider a driven-dissipative model described by the Hamiltonian (written in terms of Pauli matrices $\sigma_i^\alpha$)
  \begin{equation}\label{Eq. Hamiltonin Ising}
    H=-J \sum_{\langle i j\rangle }\sx_i \sx_j+\Delta \sum_i\sz_i\,,
  \end{equation}
 where the first term with $J>0$ \footnote{For $J<0$, we can make the transformation $\sx_i\to (-)^{|i|}\sx_i$ and $\sy \to (-)^{|i|}\sy_i$, where $|i|$ denotes the sum of the coordinates of the site $i$, that is, we make a 180-degree rotation around the $z$ axis on one of the two checkerboard sublattices. The Lindblad operator $L_i \to (-1)^{|i|} L_i$, which, however, can be gauged away; see the following discussion in the text.} is the nearest-neighbor interaction on a $d$-dimensional cubic lattice with $d=2$ or $3$, and with the dissipation via  the Lindblad operator at each site $L_i=\sqrt{\Gamma}\,\sd_i=\sqrt{\Gamma}\,(\sx_i-i\sy_i)/2$, which is simply the spontaneous emission from spin up
 $\ket{\uparrow}$ to spin down $\ket{\downarrow}$. In the absence of dissipation, the ground state of Eq.~(\ref{Eq. Hamiltonin Ising}) is known to give rise to a quantum phase transition \cite{SachdevBook} from an ordered phase (completely ordered at $J/\Delta \gg 1$), where the $\mathbb{Z}_2$ symmetry $\sigma_x \to -\sigma_x$ is broken and $\langle\sigma_x\rangle\ne 0$, to a disordered phase (fully disordered at $J/\Delta \ll 1$), where the symmetry is restored and $\langle\sigma_x\rangle=0$. For the dissipative system considered here, ${\mathbb Z}_2$ is still a symmetry of the master equation. To see this, consider the transformation
  \begin{equation}
    \sigma_x\to -\sigma_x, \quad \sigma_y\to -\sigma_y,\quad \sigma_z\to \sigma_z,
  \end{equation}
  which respects the noncommutative algebra of Pauli operators, and under which $L_i \to -L_i$. Since the master equation is bilinear in $L_i$ and $L_i^\dagger$, the overall sign (or phase) of the Lindblad operators is insignificant. One may then expect a phase transition between an ordered and a disordered phase even in the presence of dissipation. We will first briefly discuss the mean field (MF) prediction, and then compare and contrast it with the more careful Keldysh field-theoretic treatment.
  \\

  \textbf{Mean field}---For $(\z J - \Delta) \Delta  > 0$ (with $\z=2 d$ the coordination number), as $\Gamma$ is increased, the MF gives a continuous transition from the ordered phase $\langle\sx_i\rangle$ = const $\neq 0$ to a disordered phase $\langle\sx_i\rangle =0$ (see App.\ \ref{appA}). Most qualitative features of the MF are confirmed below by the field-theoretic treatment with quantitative corrections.
  One qualitative difference, however, is that the second-order transition seems {to be replaced by a first-order transition}
  for sufficiently strong dissipation.
  Furthermore, the Keldysh formalism reveals that the phase transition belongs to the universality class of the classical Ising model with an effective temperature determined by the microscopic parameters in the model. Interestingly, the effective temperature remains finite even in the limit $\Gamma\to0$.
  \\

  \textbf{Field theory}---For a proper field-theoretic treatment, we first write the spin operators in terms of hard-core bosons
  \begin{equation}\label{Eq. Hard-core bosons}
    \sd_i=a_i, \qquad \su_i=a^\dagger_i, \qquad \sz_i= 2 a^\dagger_i a_i-1\,.
  \end{equation}
  The hard-core constraint can be implemented via a large on-site potential $U\,a^\dagger_i a^\dagger_i a_i a_i$ with $U\to +\infty$. We are particularly interested in taking the continuum limit via $a_i \to a(\bx)$, where the operator $a(\bx)$ varies continuously in space. Since the MF predicts uniform phases, our assumption should be justified as a starting point for the field theory. However, as is typical with the transition to the continuum, hard-core features become soft in the continuum after coarse graining where short wavelength modes are eliminated. For example, the (near-)critical behavior of the classical Ising model with discrete values $s_i=\pm 1$ is mapped to the $\phi^4$ field theory \cite{SachdevBook,PolyakovBook}. At long wavelengths, the Ising spins can be coarse-grained to a continuous field $\phi$, and interact via a soft $\phi^4$ term which respects the original symmetry of the problem [$\sx_i \to -\sx_i$ reflected by $\phi(\bx)\to -\phi(\bx)$]; in principle, one must also include higher-order terms ($\phi^6$, etc.) that respect the symmetry, but such terms are less relevant under RG and can be dropped. The quantum TFIM also enjoys a similar mapping to the $\phi^4$ theory in the continuum albeit in one higher dimension \cite{SachdevBook}. With this in mind, we shall frequently use the mapping to the continuum, and add the quartic term introduced above with a finite strength of the interaction. The long-distance behavior of our model should be insensitive to this assumption. This is especially the case near the critical point where $\langle a \rangle$ and $\langle a^\dagger \rangle$ are small, and a phenomenological expansion and truncation of the interaction term---consistent with the underlying symmetries---at the quartic order is further justified.

  In the continuum, the first term in the Hamiltonian (\ref{Eq. Hamiltonin Ising}) written in terms of the bosonic operators should be expanded up to the second derivative in spatial coordinates; higher derivatives can be ignored in the long wave-length limit. Other terms in the Hamiltonian map to the continuum in a straightforward way. The full Hamiltonian then reads
  \begin{align}
    H=&  \int_\bx \,\, -J \,[a(\bx)+a^\dagger(\bx)]\left( d+\frac{1}{2}{ \nabla}^2\right) [a(\bx)+a^\dagger(\bx)] \nonumber \\
    &+2\Delta a^\dagger(\bx) a(\bx) + U a^\dagger(\bx)a^\dagger(\bx)a(\bx)a(\bx),
  \end{align}
  with the Lindblad operator $L_\bx= \sqrt{\Gamma} \, a(\bx)$; the lattice spacing is taken to be unity for simplicity.
  {As we shall see, this continuum model
  reproduces the MF equations for small dissipation (cf. App.\ \ref{appA}) but indicates the failure of the mean field for large dissipation.}

  Next we map the master equation to the Keldysh path integral in terms of classical and quantum fields $a_{cl,q}(t,\bx)$.
  It is more convenient to work with the real and imaginary parts of $a_{cl/q}\equiv c_{cl/q}+i d_{cl/q}$. Equations (\ref{Eq. SH}, \ref{Eq. SD}) then yield the Keldysh Lagrangian (cf. Ref.~\cite{Sieberer14})
  \begin{align}\label{Eq. TFIM}
    {\cal L}_{\rm K}&=2\left(-c_{cl} \partial_t d_q +d_{cl}\partial_t c_q \right)+ 4 J \, c_q \nabla^2 c_{cl}+ \tilde J\, c_{q} c_{cl}\nonumber \\
    -& \tilde\Delta(c_{q} c_{cl} +d_{cl} d_q)+\Gamma(c_{cl}d_q-c_q d_{cl})+i \Gamma(c_q^2+d_q^2) \nonumber \\
    -&U (c_{cl}^2+c_q^2+d_{cl}^2+d_q^2)(c_{cl} c_q+ d_{cl}d_q),
  \end{align}
  where $\tilde J= 4 \z J$, $\tilde \Delta=4\Delta$, and $\z=2d$ is the coordination number.
  Notice that there is no gradient term in the fields $d_{cl/q}$, which makes them purely local in spatial coordinates. Furthermore, they are ``gapped'' due to the terms $-\tilde\Delta d_c d_q+ i \Gamma d_q^2$ in the action (while $c_{cl/q}$ can be tuned near criticality), but we cannot just drop the $d_{cl/q}$-dependence as there would be no time derivative acting on $c_{cl/q}$.
  We thus integrate out the former, and find an effective action for the field $c_{cl/q}$. Since $d_{cl/q}$ are gapped, they can be integrated out at the level of the quadratic terms in the first and second lines of Eq.~(\ref{Eq. TFIM}); the effective Lagrangian simply follows from the saddle-point approximation as \footnote{The saddle-point approximation including the (nonlinear) interaction term gives corrections that are irrelevant in the sense of RG.}
  \begin{align}
    {\cal L}_{\rm eff}=& -\frac{4\Gamma}{\tilde\Delta} c_{q} \partial_t c_{cl}+ 4J\, c_q \nabla^2 c_{cl} + (\tilde J-\tilde\Delta-\frac{\Gamma^2}{\tilde\Delta})c_q c_{cl} \nonumber \\
    &+ i \Gamma ( 1+\frac{\Gamma^2}{\tilde\Delta^2}) c_q^2 -U (1-\frac{\Gamma^4}{\tilde\Delta^4})\,c_{cl}^3 c_q  \\
    &+{\cal O}\left(c_q \partial_t^2 c_{cl}, \,\,i (\partial_t c_q)^2, \,\, c_{cl}c_q^3\right)\,, \nonumber
  \end{align}
  where ${\cal O}(\cdot)$ contains irrelevant terms in the sense of RG due to their higher-order time derivatives, or higher powers of the quantum field (see Sec.~\ref{Sec. Framework}).
  We briefly remark that, in the absence of dissipation, the linear time derivative vanishes, and one should keep the second-order time derivative, which, combined with the second-order spatial derivatives,
  gives rise to the $\phi^4$ field theory in one higher dimension
  , and the critical point corresponds to setting
  the mass term (the coefficient of $c_{cl}c_q$ with $\Gamma\to0$) to zero. In the presence of dissipation, the linear time derivative is more relevant, and the quantum criticality is inaccessible.
 In this case, the critical point obtained by setting the mass term to zero matches exactly with the MF equation derived in App.\ \ref{appA}. However, we will see shortly that there are important caveats indicating the failure of the mean field, especially for sufficiently strong dissipation.

  Neglecting the irrelevant terms, the quantum field $c_q$ appears at most quadratically in the Keldysh action, and thus an \emph{exact} classical Langevin equation emerges for $c_{cl}$ ($c_{cl} \to c$):
  \begin{equation}\label{Eq. Langevin for c}
    -\frac{4\Gamma}{\tilde\Delta} \partial_t c=\left[-4J \,\nabla^2+r +u \,  c^2(t,\bx)\right] c(t,\bx) +\xi(t,\bx),
  \end{equation}
  with the parameters
  \begin{align}
    r=-\tilde J+\tilde\Delta+\frac{\Gamma^2}{\tilde\Delta},\qquad u=U\left(1-\frac{\Gamma^4}{\tilde\Delta^4}\right),
  \end{align}
  where the noise $\xi$ is correlated as \cite{KamenevBook,KardarBook}
  \begin{equation}
    \langle \xi(t,\bx) \xi(t',\bx')\rangle  = 2\,\Gamma\left(1+\frac{\Gamma^2}{\tilde\Delta^2}\right)\, \delta(\bx-\bx') \delta(t-t')\,.
  \end{equation}
Equation (\ref{Eq. Langevin for c}) describes dynamics that is mathematically equivalent to the dynamics near the thermodynamic equilibrium of the field $c$. In this sense, the steady-state solution to Eq.~(\ref{Eq. Langevin for c}) is simply given by the thermal Gibbs ensemble (normalizing $u$)
  \begin{equation}\label{Eq. Partition fn}
    P[c(\bx)] \sim \exp\left[-\frac{1}{T_{\rm eff}}\int_\bx \,2J(\nabla c)^2 +\frac{r}{2} \, c^2 + u c^4\right],
  \end{equation}
  with
  \begin{equation}\label{Eq. Eff temperature}
  T_{\rm eff}=\frac{\tilde\Delta}{4}\left(1+\frac{\Gamma^2}{\tilde\Delta^2}\right),
  \end{equation}
  where $T_{\rm eff}$ is obtained by imposing the fluctuation-dissipation relation discussed following Eq.~(\ref{Eq. PF}).
  We point out that the effective temperature goes to a finite value even for $\Gamma \to 0$. Therefore, even with infinitesimal dissipation, the system, contrary to what one might naively expect, does not get arbitrarily close to the ground state of the effective Hamiltonian in Eq.~(\ref{Eq. Partition fn}). At the same time, for infinitesimal dissipation, the approach to the steady state will typically take a very long time.

  The partition function defined as the sum over all configurations of $c$ weighted by the probability distribution (\ref{Eq. Partition fn}), $\int D[c]\, P[c]$, is nothing but the $\phi^4$ theory, which describes the universality class of the classical Ising model. The critical point is given by $r=0$, at which point we have $\tilde J=4\z J= \tilde\Delta(1+\Gamma^2/\tilde \Delta^2)$. Comparing against Eq.~(\ref{Eq. Eff temperature}), it becomes evident that, at the critical point, $J/T_{\rm eff}=1/\z$. Interestingly, the same relation also describes the mean-field solution for the critical point of the \emph{classical} Ising model $H=-J\sum_{\langle i j\rangle}s_i s_j$ without the transverse field and dissipation.
  This implies that the transverse field together with dissipation play the role of thermal fluctuations, where even the value of the critical temperature is matched (at, and most likely also beyond, the mean field level) with the classical Ising model. This surprising feature is most likely  related to the fact that both the transverse field and dissipation are defined in the $\sz$ basis.
  Comparing to Eq.~(\ref{Eq. Hamiltonin Ising}), one can see that the effective result of dissipation is simply to reduce the original Hamiltonian to one with only the Ising term, which, however, should be regarded at a finite effective temperature. On the other hand, in all other models that we study in this paper, the final effective Hamiltonian and the corresponding thermodynamic model bear no such obvious relationship to the original dissipative-driven model.

  For our field-theoretic treatment to be valid, the partition function should be convergent, and specifically $u>0$, i.e.\ $\Gamma <4\Delta$.
  For $\Gamma> 4\Delta$, the sign of the quartic term is negative, and the model described by Eq.~(\ref{Eq. Partition fn}) exhibits an instability towards a phase where $c\to \pm \infty$.
  Of course, the sum (trace) over spins is always convergent (spin excitations will be saturated), and thus there should be higher order terms such as $c^6$ with a positive coefficient in Eq.~(\ref{Eq. Partition fn}) making the functional-integral fully convergent. In this case, one finds that the second-order phase transition is replaced by a first-order phase transition in a model described by the effective Hamiltonian density
  ${\cal H} \sim 2J (\nabla c)^2+(r/2) c^2+ u c^4 +v c^6$ where $u<0$ and $v>0$, see p.~173 of \cite{Chaikin1995}.

  \emph{Dynamics}: The Langevin equation (\ref{Eq. Langevin for c}) clearly indicates that the dynamics is diffusive, i.e.\ the dynamic exponent is $z=2$ at the mean-field level with corrections due to fluctuations captured by loop diagrams. In short, we have reduced the starting dissipative spin system to the dynamical Landau-Ginzburg model where the dynamical field is not conserved. The latter model falls under the so-called model A dynamics of the Halperin-Hohenberg classification \cite{Hohenberg77}, where $z$ is known via epsilon-expansion or numerical evaluation.

  \subsection{Isotropic XY model with coherent drive and spontaneous emission}\label{Sec. XY with coherent drive}
  In this subsection, we consider the lattice Hamiltonian
  \begin{equation}
    H=-J \sum_{\langle i j\rangle }\sigma^+_i \sigma^-_j+h.c.\,\,+\Omega \sum_i\sigma^x_i+\Delta \sum_i \sigma^z_i\,,
  \end{equation}
  assuming nearest-neighbor interaction with $J>0$ \footnote{If $J<0$, we can still make the transformation $\sx_i\to (-)^{|i|}\sx_i$ and $\sy \to (-)^{|i|}\sy_i$. In this case, however, the coherent drive will oscillate in space rendering it irrelevant, and presumably leading to a trivial state $\left|\downarrow\downarrow\downarrow\cdots\right\rangle$.} on a $d$-dimensional cubic lattice with $d=2$ or $3$, together with the dissipative process via the Lindblad operator at each site $L_i=\sqrt{\Gamma}\,\sd_i$. Without the coherent drive, $\Omega =0$, the dissipation drives the system to a ``dark'' state where all spins are in  state $\ket{\downarrow}$; this happens because this dark state is an eigenstate of  the Hamiltonian for $\Omega=0$. To find a nontrivial steady state, we consider a finite coherent drive $\Omega\ne 0$.
  \\

  \textbf{Mean field}---The phase diagram of this model, at the level of MF, includes a region with a uniform ($\langle \sigma^\alpha_i\rangle$ = const $\neq 0$) stable phase, and a bistable regime, in which two uniform stable phases exist (see App.\ \ref{appB}). The MF predicts a critical point at which a continuous transition occurs between stable and bistable regions. While the full MF phase diagram has more structure to it, we are particularly interested in this criticality and closely investigate its vicinity. We show that the continuous phase transition is similar to that of the Ising-type liquid-gas phase transition, but find that bistability is an artifact of the MF.
  \\

  \textbf{Field theory}---We begin by writing the spin operators in terms of hard-core bosons, Eq.~(\ref{Eq. Hard-core bosons}), and implement the hard-core constraint via an on-site quartic potential as in the previous subsection. We remark that the parameters in the model can be chosen such that $\langle a\rangle$ and $\langle a^\dagger \rangle$ are small near the critical point, which further justifies the truncation of the interaction at the quartic order.
  In the anticipation of a uniform phase, we cast the Hamiltonian in the continuum as
  \begin{align}\label{Eq. Hamiltonian XY}
    H =&\int_\bx\,\, -J\,  a^\dagger(\bx) \nabla^2 a(\bx)+ \tilde\Delta \, a^\dagger(\bx) a(\bx)\nonumber \\
    &+ \Omega \left(a(\bx)+a^\dagger (\bx)\right)+U\, a^\dagger(\bx) a^\dagger(\bx) a(\bx) a(\bx),
  \end{align}
  where $\tilde \Delta=2\Delta- \z J$ with $\z$ the coordination number; we shall drop the tilde for notational convenience, but always mean $\tilde \Delta$ in the rest of this subsection.
  Similarly, the Lindblad operators go to $L_i \to \sqrt{\Gamma}\,a(\bx)$. The Keldysh action is then given by
  \begin{align}\label{Eq. Keldysh XY}
    &S_K\!\!=\!\!\int_{t,\bx}\!\!\!\!\!
    \begin{pmatrix}
      a_{cl}^* & a_q^*
    \end{pmatrix}\!\!
    \begin{pmatrix}
      0 & \!\!\!\!\!\!\!\! i\partial_t\!\!+\!J \nabla^2\!\! -\!\Delta\!\!-\!\!\frac{i\Gamma}{2} \\
      i\partial_t\!\!+\!J \nabla^2\!\! -\!\Delta\!\!+\!\!\frac{i\Gamma}{2} & i \Gamma
    \end{pmatrix}\!\!
    \begin{pmatrix}
      a_{cl}  \\
      a_q
    \end{pmatrix} \nonumber \\
    &\!\!\!-\sqrt{2}\Omega (a_q+a_q^*)-\frac{U}{2}\left(|a_{cl}|^2+|a_q|^2\right) \left(a_{cl} a_q^* + c.c.\right).
  \end{align}
  Before going further, we compare the continuum description in Eqs.~(\ref{Eq. Hamiltonian XY},\ref{Eq. Keldysh XY}) with the original lattice model. The two descriptions yield almost identical MF equations with a similar critical behavior and exhibit stable and bistable regions (MF in the continuum will be discussed shortly); however, for the corresponding regions to match exactly, $U$ is to be substituted in terms of the original parameters in the lattice model. For the sake of clarity and to avoid confusion, we shall regard the continuum description in Eqs.~(\ref{Eq. Hamiltonian XY},\ref{Eq. Keldysh XY}) as our fundamental model and a starting point for further investigation. In fact, closely related models arise naturally in the context of interacting Rydberg polaritons in free space \cite{otterbach13,Bienias14}.
 However, we maintain that, at least for the regions in the parameter space with small excitation density, the continuum description should be a good approximation to the lattice model.

  We first embark on a detailed study of the vicinity of the MF critical point.
  To this end, the mean field solution is first derived for the continuum model via $\delta S_K/\delta a_q^*=0$, yielding
  \begin{equation} \label{Eq. eqn for psi0}
    \left(-\Delta + i \frac{\Gamma}{2}\right)a_0 -\sqrt{2} \Omega -\frac{1}{2}U |a_0|^2 a_0=0\,,
  \end{equation}
  where $a_0=\langle a_{cl}\rangle$ is the MF value of the classical field.
  Depending on the parameters, this equation has either one stable solution, or three solutions, only two of which are stable. Near the critical point, these solutions are continuously connected.
  To characterize this point, we first define $\zeta\equiv -\Delta/U$, $\gamma\equiv\Gamma /U$, $o\equiv\Omega/U$, and $n\equiv |a_0|^2/2$ the excitation density (the factor of 1/2 appears due to the definition of $a_{cl}$ in terms of the original fields). Equation (\ref{Eq. eqn for psi0}) can  then be cast as
  \begin{equation}\label{Eq. eqn for n}
    \left[(\zeta-n)^2+\left(\frac{\gamma}{2}\right)^2\right]n=o^2.
  \end{equation}
    At the critical point, the above parameters are given by
   \begin{equation}\label{Eq. zetac}
    \qquad \gamma_\zeta=\frac{2\zeta}{\sqrt{3}}\,,\qquad  o_\zeta= \left(\frac{2\zeta}{3}\right)^{3/2}\!\!\!\!,
   \end{equation}
  which is easily verified by noting that the three roots of Eq.~(\ref{Eq. eqn for psi0}) become degenerate and give the critical excitation density of
  \begin{equation}\label{Eq. nc}
    n_\zeta=\frac{2\zeta}{3}\,.
  \end{equation}
  As the next step, we expand the Keldysh action, Eq.~(\ref{Eq. Keldysh XY}), around the MF solution in Eq.~(\ref{Eq. eqn for psi0}), i.e.\ $a_{cl} \to a_0+a_{cl}$, and find
  \begin{widetext}
  \begin{align}\label{Eq. K action expanded}
     {\cal S}_K=&
    \int_{\omega,\bk}A^\dagger(\omega,\bk)
    \begin{pmatrix}
         0 & D_{2\times 2}^A(\omega,\bk)  \\ \\
       D_{2\times 2}^R(\omega,\bk)  & D^K_{2\times 2}
    \end{pmatrix}
    A(\omega,\bk)\nonumber \\
    -\frac{U}{2} &\int_{t,\bx}  \left(2a_0|a_{cl}|^2+{a_0^*}\,a_{cl}^2\right)a_q^* +c.c. + \left(|a_{cl}|^2+|a_q|^2\right)\left(a_{cl}a_q^*+c.c.\right),
  \end{align}
    where $A(\omega,\bk)=\begin{pmatrix}
      a_{cl}(\omega,\bk) & a^*_{cl}(-\omega,-\bk)& a_q(\omega,\bk) & a_q^*(-\omega,-\bk)
    \end{pmatrix}^T$, and $D_{2\times 2}^{R,A,K}$ are $2\times 2$ matrices given by
  $D^K_{2\times 2}=\begin{pmatrix}
       i\Gamma & 0 \\
      0  & i\Gamma
    \end{pmatrix}$, $D_{2\times 2}^A(\omega,\bk)=\left[D^R_{2\times 2}(\omega,\bk)\right]^\dagger$, and
  \begin{equation}\label{Eq. DR}
    D^R_{2\times 2}(\omega,\bk)=
    \begin{pmatrix}
       \omega+i\Gamma/2 -J \bk^2 -\Delta -U|a_0|^2 & -(U/2)a_0^2 \\ \\
      -(U/2){a_0^*}^2  & -\omega-i\Gamma/2 -J\bk^2-\Delta-U |a_0|^2
    \end{pmatrix}.
  \end{equation}
    \end{widetext}
    The second line of Eq.~(\ref{Eq. K action expanded}) includes cubic and quartic terms in the action. To find the dissipative spectrum of fluctuations, or, more precisely, the relaxation rate, we solve $\det D_{2\times 2}^R(\omega,\bk)=0$ and find
    \begin{equation}\label{Eq. Dispersion}
      \!\omega_\bk=-i\frac{\Gamma}{2}\pm \sqrt{(\Delta+U|a_0|^2+J \bk^2)^2-(U/2)^2|a_0|^4}\,.
    \end{equation}
    It is easy to see that one of the two eigenvalues vanishes at the critical point as expected.
    To approach the critical point, we can fix $n=n_\zeta$ (or equivalently fix $a_0$) and $o=o_\zeta$ according to Eqs.~(\ref{Eq. zetac},\ref{Eq. nc}), and take $\gamma \to \gamma_\zeta$. For $\gamma> \gamma_\zeta$, there is a unique solution to Eqs.~(\ref{Eq. eqn for psi0},\ref{Eq. eqn for n}), while, for $\gamma<\gamma_\zeta$, two stable solutions continuously emerge. Casting $\omega$ in units of $U$, and working in the long wave-length limit, we find the relaxation rates as
    \begin{equation}
     \omega_\bk\approx \left\{
       -i\left(\frac{\gamma}{2}- \frac{\zeta}{\sqrt{3}}\right)-i\bk^2\,,\,\,\, \\
       -i\left(\frac{\gamma}{2}+\frac{\zeta}{\sqrt{3}}\right)\right\},
    \end{equation}
    where the coefficient of $\bk^2$ in the first eigenvalue is proportional to $J$ but is normalized to unity (by rescaling space), while, in the second eigenvalue, the momentum dependence is entirely neglected due to the dissipative gap.
    We are further interested in finding
        {a natural basis for} $D^R_{2\times 2}(\omega,\bk)$
    {in order} to break the fluctuations of the field into massless and massive components. To this end, we first drop the $\bk$-dependence in Eq.~(\ref{Eq. DR}) as we are interested in the long-wavelength limit; we will deal with the $\bk^2$ term later. (On similar grounds, we can also drop $\omega$, but it does not further simplify our task.) Also, to simplify computations, we absorb the phase of $a_0$ in the definition of $a_{cl/q}$, thus $a_0, a_0^* \to |a_0|$ in the off-diagonal elements of the matrix in Eq.~(\ref{Eq. DR}) as well as the coefficient of the cubic term in Eq.~(\ref{Eq. K action expanded}). The resulting matrix $D^R_{2\times 2}$ takes the form (in units of $U$)
   \begin{equation}
    D^R_{2\times 2}(\omega,\bk)\to
        \begin{pmatrix}
       \omega+i\gamma/2 -\zeta/3 &  -2\zeta/3 \\ \\
      -2\zeta/3  & \hskip -.18in-\omega-i\gamma/2 -\zeta/3
    \end{pmatrix}.
   \end{equation}
   To simplify the form of the action, we change the basis to
   \begin{align}\label{Eq. eigenbasis}
    c_{cl/q}(\omega,\bk)&=\mp \left[e^{\pm i\pi/6}a_{cl/q}(\omega,\bk)+e^{\mp i\pi/6}a_{cl/q}^*(-\omega,-\bk)\right],\nonumber \\
    d_{cl/q}(\omega,\bk)&=e^{\mp i\pi/6}a_{cl/q}(\omega,\bk)+e^{\pm i\pi/6}a_{cl/q}^*(-\omega,-\bk),
  \end{align}
  which allows us to rewrite the quadratic part of the Keldysh action as
  (for the moment, not including the $\bk$ dependence in the kernel and the quadratic quantum-noise term)
  \begin{align}\label{Eq. Keldysh 2 diagonal}
    {\cal L}_K^{(2)}\sim \,\,\, &c_q(-\omega,-\bk)c_{cl}(\omega,\bk)\left[i\omega - \frac{\gamma}{2}+\frac{\zeta}{\sqrt{3}}\right] \nonumber \\
    +&\,d_q(-\omega,-\bk)d_{cl}(\omega,\bk)\left[i\omega - \frac{\gamma}{2}-\frac{\zeta}{\sqrt{3}}\right],
  \end{align}
  where we have neglected a multiplicative constant. Equation (\ref{Eq. Keldysh 2 diagonal}) clearly indicates that $c_{cl/q}$ can be tuned near criticality by taking $\gamma \to \gamma_\zeta$, while $d_{cl/q}$ are always gapped.
  Also notice that it follows from Eq.~(\ref{Eq. eigenbasis}) that the fields $c_{cl/q}(t,\bx)$ and $d_{cl/q}(t,\bx)$ are real-valued.
  Inverting the basis in Eq.~(\ref{Eq. eigenbasis}), and casting it in real space and time, we find
  \begin{align}
    a_{cl/q}(t,\bx)=\frac{1}{\sqrt{3}}\left(\mp e^{\mp i\pi/3}c_{cl/q}(t,\bx)+e^{\pm i\pi/3}d_{cl/q}(t,\bx)\right).
  \end{align}
  With this representation, we can now cast various terms in the action in terms of the new fields. The gradient term $a_q^* \nabla^2 a_{cl}+c.c.$, i.e. the $\bk^2$ term in the action, now takes the form (with a normalized coefficient)
   \begin{equation} \nonumber
     c_q \nabla^2 c_{cl} + \cdots ,
   \end{equation}
   where the ellipses denote $d_{cl/q}$-dependent gradient terms, which are ignored since $d_{cl/q}$ are gapped \footnote{There are also cross terms such as $c_q \nabla^2 d_{cl}+d_q \nabla^2 c_{cl}$ which, upon integrating out $d$, give rise to $\sim (\nabla^2  c_{q})^2+ c_q (\nabla^2)^2 c_{cl}$, and are thus irrelevant in the RG sense.}.
   Next we cast the interaction terms in Eq.~(\ref{Eq. K action expanded}) in the new basis. We start with the cubic interaction,
   \begin{equation}\nonumber
     \lambda \left[c_{cl}^2 d_q +d_{cl}^2 c_q\right].
   \end{equation}
   The precise value of the coefficient $\lambda$
   will not be important.
   Next, the quartic term expanded in terms of $c_{cl/q}$ and $d_{cl/q}$ generates many terms (with $u>0$):
   \begin{align}
     -u &\left(c_{cl}^2+c_q^2+d_{cl}^2+d_q^2+c_{cl}d_{cl}-c_q d_q\right) \nonumber \\ &\times\left( c_{cl}c_q-d_{cl}d_q+2c_q d_{cl}-2c_{cl}d_q\right). \nonumber 
   \end{align}
   And finally the noise term (in units of $U$) takes a simple form in the new basis $i\gamma |\psi_q|^2\sim i \gamma \left(c_q^2+d_q^2-c_qd_q\right)$.
   In writing the full Keldsyh action, we note that the scaling dimension of the fields renders terms with two or more powers of the quantum field $c_q$ irrelevant (except the noise term), as explained in Sec. \ref{Sec. Framework}. We keep the relevant terms in the field $c_{cl/q}$, but also include some (not all) cross terms with $d_{cl/q}$ for reasons that will become clear shortly.
   The Keldysh action then reads
   \begin{align}\label{Eq. Full K action}
     {\cal L}_K \approx &\,\,
      c_q (-\partial_t -r +\nabla^2)c_{cl}-u \,c_{cl}^3 c_q +i \gamma \, c_q^2 \nonumber \\
     +& \,d_q(-\partial_t -R)d_{cl} +i \gamma \,d_q^2  \nonumber \\
      +&\,\lambda \left(c_{cl}^2 d_q +d_{cl}^2 c_q\right) -2u \,c_{q}d_{cl}^3+3u \,c_{cl}^2d_{cl}d_q+\cdots \nonumber\\
      -&\, i\,\gamma \,c_q d_q\,,
   \end{align}
   where
   \begin{equation}
     r=\frac{\gamma}{2}-\frac{\zeta}{\sqrt{3}}\,, \qquad R=\frac{\gamma}{2}+\frac{\zeta}{\sqrt{3}}\,.
   \end{equation}
   We have organized Eq.~(\ref{Eq. Full K action}) into (the first line) terms depending on $c_{cl/q}$ only, (the second line) quadratic terms in $d_{cl/q}$, and (the third and fourth lines) various $d_{cl/q}$-dependent nonlinear and cross terms; the ellipses denote nonlinear terms not written explicitly. Also we have not fully kept track of various coefficients (except those of $r$ and $R$) since they will be inconsequential for our conclusions.

   If we simply drop the $d_{cl/q}$-dependent terms, we get for the Keldysh action the first line of Eq.~(\ref{Eq. Full K action}). By taking the second step of writing the corresponding Langevin equation and subsequently mapping to the partition function
   {as outlined in Sec.~\ref{Sec. Framework}} (see also the previous subsection), we would find the thermodynamic Landau-Ginzburg $\phi^4$ theory with the ${\mathbb Z}_2$ symmetry, and the associated second-order phase transition as $r \to 0$.
   Taking into account the fields $d_{cl/q}$ and their fluctuations, we next show that the $\mathbb{Z}_2$ symmetry is spoiled, but a continuous phase transition emerges akin to that of the liquid-gas transition.
   To start with, let us consider the effect of fluctuations to first order in $\lambda$; at this order, we find
   \begin{equation}\label{Eq. 1st order fluc}
     \lambda \left\langle d_{cl}^2\right\rangle c_q,
   \end{equation}
   generated by integrating out $d_{cl}$ in the cubic term. This term acts like a ``magnetic'' field, and breaks the ${\mathbb Z}_2$ symmetry of $c_{cl,q}\to -c_{cl,q}$. However, it can be absorbed into the parameter $\Omega$ in the original model.
   In fact, fluctuations can modify the position of the critical point, and thus the above equation can be regarded as the correction to the MF position of the phase transition.
   We thus consider the effect of fluctuations at higher orders in $\lambda$ and $u$. The nonlinear terms in (the third line of) Eq.~(\ref{Eq. Full K action}) expanded to  second order generate various terms; of particular interest to us is the term proportional to
   \begin{align}\label{Eq. eff action to 2nd order}
     \lambda u\, \Big[c_{cl}^2 c_q\Big]_{t,\bx} \int d\tau  \,\Big\{
     \,\,& 2\left\langle d_{cl}^3(t)d_{q}(t+\tau)\right\rangle \nonumber \\
     -&3\left\langle d_{cl}(t) d_q(t)d_{cl}^2(t+\tau)\right\rangle  \,\Big\}\nonumber \\
     \sim
     \,\lambda u\, \Big[c_{cl}^2 c_q\Big]_{t,\bx} \int d\tau  \,G^{R}&(\tau)\left[G^{K}(0)-G^{K}(\tau)\right],
   \end{align}
   with all the fields evaluated at the same spatial coordinates since there is no gradient term in $d_{cl/q}$. Also $G^{R}(\tau)=-i\left\langle d_{cl}(t)d_q(t+\tau)\right\rangle$ and $G^{K}(\tau)=-i\left\langle d_{cl}(t)d_{cl}(t+\tau)\right\rangle$ are the retarded and Keldysh Green's functions, respectively, for the fields $d_{cl/q}$, which in Fourier space become $G^R(\omega)\sim (i\omega-R)^{-1}$ and $iG^K(\omega)\sim G^R(\omega)\,\gamma\, {{G^R}(\omega)}^{\!*}$. These functions have a nonzero support only for $\tau \lesssim 1/{R} \sim \gamma^{-1}$, which is why the vertex in $c_{cl/q}$ in Eq.~(\ref{Eq. eff action to 2nd order}) is approximated to be local in time. Due to the nonvanishing integral in this equation, fluctuations will generate a term of the form $v\, c_{cl}^2 c_q$ in the action with some constant $v$.
   With the first- and cubic-order terms generated as the result of fluctuations, there is no apparent ${\mathbb Z}_2$ symmetry; the full partition function then takes the form ($c_{cl} \to c$)
   \begin{equation}\label{Eq. Partition fn c}
     \int D[c(\bx)] \exp\left[-\frac{1}{\gamma}\int_\bx(\nabla c)^2+h c+r c^2 +vc^3+ u c^4\right],
   \end{equation}
   where the exact coefficients of various terms are disregarded \footnote{Higher-order odd terms (fifth order and higher) become increasingly more irrelevant under RG, and can be neglected. Nevertheless, they will give corrections to the linear- and the third-order terms, which can always be absorbed in the coefficients of the corresponding terms.}. Similar steps to those
   {outlined in Sec.~\ref{Sec. Framework}} (see also Sec.~\ref{Sec. TFIM}) are taken to obtain the partition function from the Keldysh action and the resulting Langevin equation.
   Terms with odd powers of $c$ should be traced back to the fact that the full action (\ref{Eq. K action expanded}) as a function(al) of $\{c_{cl,q},d_{cl,q}\}$ is not symmetric under the simultaneous transformations $c_{cl,q}\to -c_{cl,q}$ and $d_{cl,q}\to -d_{cl,q}$.
   Despite this fact, we can absorb the linear term in the parameter $\Omega$, and shift the field $c$ by a constant ($c \to c_0+c$) to eliminate the third-order term; the constant term merely modifies the MF critical point around which we have expanded the action. A similar scenario arises in the liquid-gas phase transition, where there is no obvious symmetry, however, one can choose parameters such as density to eliminate odd terms. This phase transition, despite the absence of symmetry, is of the Ising type \cite{Chaikin1995}.
   We thus conclude that the driven-dissipative model considered in this subsection undergoes a continuous Ising-type phase transition.
   We further remark that there is no true thermodynamic bistability; the true steady state of the system is given by the minimum of the exponent in the partition function (\ref{Eq. Partition fn c}), which is unique for generic values of $h$ and $v$. We stress that our argument does not rely on a similar minimum criterion in thermodynamics, but is simply derived in our nonequilibrium model; in the thermodynamic limit, a minimum of the exponent [in Eq.\ (\ref{Eq. Partition fn c})], being extensive in the system size, is infinitely more likely to occur than any other state including other local minima of the exponent.

   Similar models describing a dissipative gas of Rydberg atoms have been studied in Refs.~\cite{Lee12,Marcuzzi14}. A notable difference is that the interaction in Refs.~\cite{Lee12,Marcuzzi14} is a long-range interaction of the Ising $\sz_i \sz_j$ type (in contrast to the nearest-neighbor flip-flop interaction in the present subsection).
   Nevertheless, the corresponding MF analysis performed in Ref.~\cite{Marcuzzi14} is almost identical to the MF equation of this subsection. Therefore, at least in uniform phases, the model in Ref.~\cite{Marcuzzi14} might be amenable to a similar treatment. It would be interesting to study the effect of fluctuations beyond the mean field; see also the comparison with an approach based on a variational principle for steady states \cite{Weimer15}. Finally, we note that our considerations here should be directly applicable to uniform phases of driven-dissipative Bose-Hubbard models and their critical behavior studied in Refs.~\cite{Ciuti13,biella14}.

  \subsection{Anisotropic XY model with spontaneous emission}\label{Sec. Anisotropic XY}

  In this subsection, we consider the Hamiltonian
  \begin{equation}\label{Eq. Hamiltonian XYZ}
    H=\frac{J}{2d} \sum_{\langle i j\rangle } \,\, \sx_i \sx_j- \sy_i \sy_j =\frac{J}{d} \sum_{\langle i j\rangle } \,\, \su_i \su_j + \sd_i \sd_j\,,
  \end{equation}
   assuming nearest-neighbor interactions with $J>0$ on a $d$-dimensional cubic lattice with $d=2$ or $3$, together with dissipation via the Lindblad operator at each site $L_i=\sqrt{\Gamma}\,\sd_i$.  While  spontaneous emission tends to create a state where all spins point down,
 such a state is not an eigenstate of the Hamiltonian. The interplay of the dissipation and the effective drive in the Hamiltonian can give rise to nontrivial steady states. An experimental realization of this model using ultracold atoms in the ground electronic state weakly dressed with highly excited Rydberg states is proposed in Ref.\  \cite{Lee13}.
   \\

   \textbf{Mean field}---The MF is studied in Ref.\ \cite{Lee13} for the more general XYZ model with $J_x\ne J_y\ne J_z$. For sufficiently weak spontaneous emission in the model considered here, the MF (see App.\ \ref{appC}) predicts a spontaneous symmetry breaking that gives rise to a staggered XY steady state with spins on neighboring sites pointing in different directions. For larger values of spontaneous emission, one finds a disordered paramagnetic state. Our field-theoretic treatment confirms this picture in three dimensions, however, in two dimensions, we show that the XY phase cannot be realized. This happens because the effective temperature, emerging due to the dissipation, is larger that the Kosterlitz-Thouless temperature associated with the transition from short-range to algebraic long-range order in two dimensions.
   \\

   \textbf{Field theory}---As in the previous subsections, we first map spins to hard-core bosons, and represent their hard-core nature via a quartic term. This mapping is particularly good close to the phase transition between the paramagnetic and the XY phase because $\sz\approx -1$ near the phase boundary. In  anticipation of the staggered XY phase, we recast the Hamiltonian on two checkerboard sublattices $A$ and $B$ as
   \begin{equation}
     H= \frac{\tilde J}{2} \sum_{\langle ij\rangle}( b^\dagger_i a^\dagger_j+b_i a_j )+ U\sum_{i\in A} a^\dagger_i a^\dagger_i a_i a_i+ U\sum_{j\in B} b^\dagger_i b^\dagger_i b_i b_i\,,
   \end{equation}
   where $\tilde J\equiv 2J/d$. Note that we have chosen the same coefficient for the interaction terms on both sublattices to make manifest the symmetry of the underlying spin model.
   The Keldysh Lagrangian at the quadratic level, and in Fourier space, can be written as (the integral over frequency is limited to $\omega>0$)
   \begin{widetext}
   \begin{align}
     {\cal L}_K^{(2)}=&
     \begin{pmatrix}
      a_{cl}^* & a_q^*
    \end{pmatrix}_{\omega,\bk}
    \begin{pmatrix}
      0 & \omega-i\frac{\Gamma}{2} \\
      \omega +i\frac{\Gamma}{2} & i \Gamma
    \end{pmatrix}
    \begin{pmatrix}
      a_{cl}  \\
      a_q
    \end{pmatrix}_{\omega,\bk}
    \,\,     +     \,\,\,
    \begin{pmatrix}
      b_{cl}^* & b_q^*
    \end{pmatrix}_{\omega,\bk}
    \begin{pmatrix}
      0 & \omega-i\frac{\Gamma}{2} \\
      \omega +i\frac{\Gamma}{2} & i \Gamma
    \end{pmatrix}
    \begin{pmatrix}
      b_{cl}  \\
      b_q
    \end{pmatrix}_{\omega,\bk} \nonumber \\
    &+\tilde J(\bk) \left[a_q^*(-\omega,-\bk) b_{cl}^*(\omega,\bk)+a_{cl}(-\omega,-\bk) b_q(\omega,\bk)+c.c.\right],
   \end{align}
   \end{widetext}
   where we have defined $\tilde J(\bk)= \tilde J \left[\cos(k_x)+\cos(k_y)+\cdots\right]$, with dots standing for $\cos(k_z)$ in the case of $d=3$ dimensions.
   A quadratic Keldysh action with the assumption that spins map to soft-core bosons, and an ansatz of a uniform phase (i.e., by identifying $a$ and $b$ fields), has been first constructed in Ref.~\cite{Lee13}.
   The above quadratic action becomes diagonal in the basis defined by
   \begin{align}\label{Eq. basis XYZ}
     c_{cl/q}(\omega, \bk)&=\mp \left[e^{\pm i\pi/4} b_{cl/q}(\omega, \bk)+e^{\mp i\pi/4} a_{cl/q}^*(-\omega,-\bk)\right], \nonumber \\
     d_{cl/q}(\omega, \bk)&=e^{\mp i\pi/4} b_{cl/q}(\omega, \bk)+e^{\pm i\pi/4} a_{cl/q}^*(-\omega,-\bk),
   \end{align}
   which casts the quadratic part of the Keldsyh Lagrangian into
   \begin{align}\label{Eq. Keldsyh 2 Lag}
     {\cal L}_K^{(2)}= \frac{1}{2}\, \Big\{\, &\big[i \omega-\frac{\Gamma}{2}+ \tilde J(\bk)\big]c_q^*(\omega,\bk) c_{cl}(\omega,\bk) +c.c. \nonumber \\
     +&\big[i \omega-\frac{\Gamma}{2}- \tilde J(\bk)\big]d_q^*(\omega,\bk) d_{cl}(\omega,\bk)+c.c. \nonumber \\
      +&\,\, i \Gamma \left(|c_q(\omega,\bk)|^2+|d_q(\omega,\bk)|^2\right)\Big\}.
   \end{align}
   At long wavelengths, $\bk \to 0$, and $\tilde J (\bk)\to \tilde J({\bf 0})=2J>0$. Therefore, the fields $c_{cl/q}$ can be tuned near criticality with a vanishing gap characterized by $r=\Gamma/2-2J$, while $d_{cl/q}$ are massive with (the imaginary part of) the gap $R=\Gamma/2+2J$, and can be integrated out; however, we must not drop them before considering the (nonlinear) interaction terms.
   Casting the quartic interaction terms in the continuum and then in the Keldysh basis, we get $(-U/2)\int_{t,\bx} \left(|a_{cl}|^2+|a_q|^2\right) \left(a_{cl} a_q^* + c.c.\right)$ plus a similar term for $b_{cl/q}$. To write the interaction in the new basis, we invert the eigenbasis in Eq.~(\ref{Eq. basis XYZ}) and cast it in space-time coordinates,
   \begin{align}
     a_{cl/q}(t,\bx)&=\frac{1}{2}\left[\mp e^{\mp i\pi/4}c^*_{cl/q}(t,\bx)+e^{\pm i\pi/4}d^*_{cl/q}(t,\bx)\right],\nonumber \\
     b_{cl/q}(t,\bx)&=\frac{1}{2}\left[\mp e^{\mp i\pi/4}c_{cl/q}(t,\bx)+e^{\pm i\pi/4}d_{cl/q}(t,\bx)\right].
   \end{align}
   The interaction then takes the form (up to a multiplicative constant)
   \begin{align}
     U\int_{t,\bx}& \left(|c_{cl}|^2+|d_{cl}|^2+|c_q|^2+|d_q|^2\right)\left(-c_{cl} d_q^*+d_{cl} c_q^*+c.c.\right) \nonumber \\
     &-\left(c_{cl} d_{cl}^*+c_q d_q^*-c.c.\right)\left(c_{cl}c_q^* +d_{cl}d_q^*-c.c.\right). \label{Eq. Interaction in c-d 2}
   \end{align}
   If we simply drop $d_{cl/q}$, there will not be any nonlinear terms. Instead we should find the new interaction vertices generated via integrating out $d_{cl/q}$. We thus expand the Keldysh functional integral to the first few orders in the interaction. To  first order, the effective interaction vanishes once averaged by the Gaussian functional integral due to Eq.~(\ref{Eq. Keldsyh 2 Lag}) since all the terms in Eq.~(\ref{Eq. Interaction in c-d 2}) are odd in $d_{cl/q}$. To  second order, we generate the vertex (up to a positive prefactor \footnote{Contraction of various terms within the second-order perturbation theory in the expression (\ref{Eq. Interaction in c-d 2}) produces various terms. The coefficient of the term $c_q^* c^*_{cl}c^2_{cl}$ in the effective action is given by
   \[
   \frac{1}{i}(-iU)^2\!\!\int \!\!\!d\tau (iG^K(0)) \left(-48iG^R(\tau) +16 iG^A(\tau)\right),
   \] which produces Eq.~(\ref{Eq. Eff vertex}) up to a factor of 32.})
   \begin{align}\label{Eq. Eff vertex}
     U^2 \, \Big[|c_{cl}|^2  c_{cl} c_q^*\Big]_{t,\bx} \,\,iG^{K}(0)\!\!\int \!\!d\tau  \,G^{R}(\tau)\,.
   \end{align}
   Many terms contribute to this vertex, but its precise coefficient is immaterial for our considerations.
   As in the previous subsection, the fields are evaluated at the same spatial coordinates (the gradient term in $d_{cl/q}$ is ignored due to the dissipative gap). Also the retarded and Keldysh Green's functions $G^{R/K}$ for the fields $d_{cl/q}$ are, similar to the expressions given before, $G^R(\omega)\sim [i\omega- R]^{-1}$ and $iG^K(\omega)\sim G^R(\omega)\,\Gamma\, {{G^R}(\omega)}^{\!*}$ with a nonzero support for $\tau \lesssim \Gamma^{-1}$, hence the local form of Eq.~(\ref{Eq. Eff vertex}) in time $t$. In short, fluctuations will generate a term of the form $-u\, |c_{cl}|^2 c_{cl} c_q^*$ in the action; the facts that $\int d\tau G^R(\tau)=G^R(\omega=0)\sim -1/R<0$ and $iG^K(\tau=0)\sim \Gamma/R>0$ ensure that $u>0$. Various other terms generated by integrating out $d_{cl/q}$ either give corrections to the existing terms in the action, or produce irrelevant terms in the sense of RG. Importantly, the effective action respects the $U(1)$ symmetry ($c\to c\, e^{i \theta}$) beyond the quadratic order.
   The final form of the Keldysh Lagrangian, with the relevant terms only, becomes
   \begin{align}
     {\cal L}_K =&\frac{1}{2} \Big\{ \, c_q^* \big(-\partial_t+\frac{1}{2}\tilde J\,\nabla^2 -r \big) c_{cl}+c.c. \nonumber \\
     &-u |c_{cl}|^2 \left( c_{cl} c_q^*+c.c. \right) +i\Gamma |c_q|^2\Big\}.
   \end{align}
   The quantum vertex appears at most quadratically, leading to a classical Langevin equation with a noise term which can be interpreted as an effective temperature. The corresponding steady state is then described by the thermodynamic partition function ($c_{cl}\to \psi$)
   \begin{equation} \label{Eq. XY partition function}
     \int D[\psi (\bx)]\exp\left[-\frac{1}{T_{\rm eff}}\int_\bx \frac{1}{2}\,\tilde J\,|\nabla \psi|^2+r\, |\psi|^2 +u |\psi|^4\right],
   \end{equation}
   where $T_{\rm eff}=\Gamma+\cdots$ with the ellipses being the corrections due to renormalization; the effective temperature and its precise coefficient (unity) are obtained using the fluctuation-dissipation condition. The partition function (\ref{Eq. XY partition function}) belongs to the universality class of the classical XY model.
   One should then expect off-diagonal order in $d=3$ dimensions, and a Kosterlitz-Thouless (KT) transition in $d=2$ dimensions.
   {Nevertheless, we remark that a possible emergence of the Kardar-Parisi-Zhang equation \cite{Kardar86} similar to Ref.~\cite{Altman15} may invalidate the above analysis in two dimensions; however, we show that, even in the absence of such mechanism, the constraints on the Kosterlitz-Thouless temperature would prevent the system from realizing the XY phase in two dimensions.
   To see this, }we should examine the condition for the KT transition. Denoting $\psi =|\psi| e^{i \theta}$ in the ordered phase, the algebraic long-range order is realized when ($\tilde J=J$ in $d=2$ dimensions) \cite{KardarBook}
   \begin{equation} \label{Eq. BKT condition}
     \frac{ J |\psi|^2}{T_{\rm eff}} >\frac{2}{\pi}\,.
   \end{equation}
   Normally, at sufficiently low temperatures, this condition is well satisfied; however, in this case, as $T_{\rm eff} \to 0$, or equivalently $\Gamma \to 0$, we also find---from the mean-field analysis---that $\psi \to 0$ in the same limit. To see this, note that
   \begin{equation}
   \psi \to c = e^{i\pi/4}a+e^{-i\pi/4} b^\dagger \to
   \frac{1}{\sqrt{2}}(\sx+\sy).
   \end{equation}
   Therefore, $|\psi|^2\equiv |\langle c\rangle|^2=2 \langle \sx\rangle^2$ where we have taken $\langle \sx\rangle=\langle \sy\rangle$; the value of $\langle \sx\rangle $ is inserted from the mean-field analysis in two dimensions,
   \begin{equation}
     \langle \sx\rangle =\frac{\sqrt{4 J \Gamma -\Gamma^2}}{4J}\,.
   \end{equation}
   With these expressions, and $T_{\rm eff}\approx\Gamma$, the condition (\ref{Eq. BKT condition}) takes the form
   \begin{equation}
     \frac{1}{4}-\frac{1}{16 j}>\frac{1}{\pi}\,,
   \end{equation}
   with $j\equiv {J}/{\Gamma}$ being larger than $1/4$ in the XY phase, a condition that is never satisfied, and becomes even worse with increasing $J$ (or decreasing $\Gamma$). Of course, in evaluating the above expressions, we have used mean-field expressions which can be modified, and relied on our field theory description away from phase boundaries. However, one should expect that the algebraic long-range order in two dimensions will be significantly diminished, if not completely disappear.

\emph{Dynamics}: In $d=3$ dimensions, the Langevin equation corresponding to Eq.~(\ref{Eq. XY partition function}) indicates that the dynamics is diffusive. The dynamical field is not conserved, and thus the dynamics falls under Model A of the Hohenberg-Halperin classification for the Landau-Ginzburg model with $N=2$ components \cite{Hohenberg77}.

   \subsection{Isotropic XY model with incoherent pumping and interaction-induced loss}\label{Sec. XY with two-body loss}

  In this subsection, we consider the Hamiltonian
  \begin{equation}
    H=-J \sum_{\langle i j\rangle }\sigma^+_i \sigma^-_j +h.c.\,\,+ \Delta \sum_i \sigma^z_i\,,
  \end{equation}
  with $J>0$ on a three-dimensional cubic lattice \footnote{The two-dimensional case requires a more careful treatment, see Ref.~\cite{Altman15}.}.
  In all the models in the previous subsections, we have chosen a simple dissipative process where spins at different sites spontaneously and independently decay from $\ket{\uparrow}$ to $\ket{\downarrow}$. In this subsection, in addition to the spontaneous emission via $L^l_i=\sqrt{\Gamma}\,\sd_i$,
  we also consider pumping defined via $L^p_i=\sqrt{\Gamma_p}\,\su_i$, and an interaction-induced loss described by the Lindblad operator $L^I_{ij}=\sqrt{\kappa} \, \su_j \sd_j \sd_i$ for nearest neighbors $i$ and $j$.
  The latter is an example of a more complicated type of dissipation that depends on the correlation between nearby sites. In this case, the jump operator $L^I$ checks if there are two excitations on neighboring sites, and, if there are, kills one. This type of operator is natural in systems where a particular laser coupling scheme creates dark states, or pseudo-spin states, as linear combinations of the microscopic energy levels; however, interaction between neighboring pseudo-spin states shifts them out of resonance, and can lead to the decay of one of them \cite{gorshkov11,Peyronel12,gorshkov13b}.
  \\

  \textbf{Mean field}---Without the interaction-based loss, this model is rather trivial. For a fixed decay rate $\Gamma$, the excitation density increases with increasing $\Gamma_p$, but $\langle \sx\rangle=\langle \sy\rangle=0$ since the steady state can be easily seen to be a product of single-site density matrices diagonal in the $\sigma_z$ basis.
  Specifically, for $\Gamma_p=\Gamma$, the system is in an infinite-temperature state.   At the level of the MF, the same qualitative behavior persists even in the \emph{presence} of the interaction-induced loss, although it changes quantitatively.
  More importantly, the phase predicted by the MF does not break the $U(1)$ symmetry of the problem, i.e. $\langle \sx\rangle=\langle \sy\rangle=0$; see App.\ \ref{appD}. However, we shall see that the Keldysh path-integral approach gives a continuous transition from the disordered phase to a phase with spontaneous continuous symmetry breaking.
  \\

  \textbf{Field theory}---We start by assuming that $\kappa \gg \Gamma,\Gamma_p$. With this assumption, the excitation density is rather small, and one can safely represent the spins in terms of soft-core bosons but with quartic interactions as in the previous subsections. We then find the Lindblad operator  $L^I_{ij}=a^\dagger_j a_j a_i$. In the continuum description, this operator can be cast as $L^I \to a^\dagger(\bx)a^2(\bx)$ plus gradient terms which are less relevant \footnote{{In general, non-commuting terms in the Lindblad operator require special care in mapping to the Keldysh action.
  A more careful treatment may give rise to one- and two-body loss terms, which nevertheless would not change the conclusions of this subsection.}}. Some algebra yields the Keldysh Lagrangian (with a normalized $J$)
  \begin{widetext}
  \begin{align}
     {\cal L}_K=&
    \begin{pmatrix}
      a_{cl}^* & a_q^*
    \end{pmatrix}
    \begin{pmatrix}
      0 & i\partial_t+J \nabla^2 +\mu-i\frac{\Gamma-\Gamma_p}{2} \\ \\
      i\partial_t +J \nabla^2+\mu+i\frac{\Gamma-\Gamma_p}{2} & i (\Gamma+\Gamma_p)
    \end{pmatrix}
    \begin{pmatrix}
      a_{cl}  \\
      a_q
    \end{pmatrix} \cr
    &-U |a_{cl}|^2 \left(a_{cl} a_q^* + c.c.\right)-i\frac{ \kappa}{2} |a_{cl}|^4 \left(a_{cl}^* a_q-c.c.\right),
  \end{align}
  \end{widetext}
  where we have introduced $\mu$ using the freedom (due to the symmetry) in choosing a rotating frame with $\langle a \rangle \sim \exp(-i\omega_0t)$. In writing the Lagrangian, we have ignored terms at the quadratic or higher orders in the quantum field---except the noise term $i |a_q|^2$---as they are irrelevant in the sense of RG. Casting the Keldysh path integral as a Langevin equation, we find ($a_{cl}\to \psi$)
  \begin{align}\label{Eq. Langevin pump}
    &\Big[i\partial_t +J \nabla^2 +\mu+ i \frac{\Gamma-\Gamma_p}{2}-U |\psi|^2+i\frac{\kappa}{2}|\psi|^4\Big]\psi(t,\bx) \nonumber \\
    &=\,\xi(t,\bx)\,,
  \end{align}
  with
  \begin{equation}
    \langle \xi(t,\bx) \xi^*(t',\bx') \rangle =(\Gamma+\Gamma_p)\delta(t-t') \delta(\bx-\bx')\,.
  \end{equation}
  Before considering fluctuations, we study the mean field solution at the level of the Keldysh action or the corresponding Langevin equation (\ref{Eq. Langevin pump}). We shall see that even the mean field, at the level of the path integral, improves upon the MF on the lattice model.
  In the absence of fluctuations, a uniform phase exists provided that $\Gamma_p> \Gamma$,
  \begin{equation}\label{Eq. theta}
    \psi= \left(\frac{\Gamma_p-\Gamma}{\kappa}\right)^{1/4} \, e^{i\theta},
  \end{equation}
  for a constant phase $\theta$; the real part of the bracket in Eq.~(\ref{Eq. Langevin pump}) vanishes by appropriately choosing the constant $\mu$. The solution in Eq.~(\ref{Eq. theta}) explicitly breaks the $U(1)$ symmetry.
  In fact, a similar model was studied in Refs.~\cite{Sieberer13,Sieberer14} where the authors concluded that the system becomes purely dissipative under RG, and specifically the real part of the coefficients of the gradient and nonlinear terms vanishes in the long wave-length limit.
  Our model is slightly different because the dissipative nonlinearity arises at the fifth, rather than the third, order in Eq.~(\ref{Eq. Langevin pump}); however, the RG procedure (and, most intuitively, momentum-shell RG) creates all possible terms consistent with symmetry. Therefore, our model flows to the same universality class as the model in Refs.~\cite{Sieberer13,Sieberer14}, or, equivalently, that of Eq.~(\ref{Eq. XY partition function}) with $r= (\Gamma-\Gamma_p)/2$ and $T_{\rm eff}=\Gamma+\Gamma_p$. Specifically, the gradient term, being purely coherent at the level of the original Hamiltonian, becomes dissipative in the course of RG.

  Similar considerations apply to a different model described by the Lindblad operator $L^I_{ij}\sim \sd_j \sd_i$ which kills \emph{both} excitations on neighboring sites. The lattice MF analysis again fails to capture the full phase diagram. The field-theoretic treatment in this case becomes almost identical to the model in Refs.~\cite{Sieberer13,Sieberer14}  which then predicts a phase with spontaneous continuous symmetry breaking.

\section{Conclusions and outlook}
In this paper, we have studied nonequilibrium steady states of a number of driven-dissipative systems
exhibiting a nontrivial competition between drive and dissipation.
{Mean field theory is often used to predict the many-body phases and phase transitions of such systems. However, a more careful field-theoretic treatment based on the Keldysh formalism may invalidate certain predictions of mean field analysis.
We saw that, for example, bistability is an artifact of the mean field theory (model in Sec.~\ref{Sec.   XY with coherent drive}).}
Sufficiently strong dissipation may also make certain phases inaccessible (model in Sec.~\ref{Sec.   Anisotropic XY} in $d=2$ dimensions), or may turn a continuous transition into a first-order phase transition (model in Sec.~\ref{Sec. TFIM}).
More generally, the path-integral approach and even its classical (saddle-point) approximation produces better results than mean field theory not only in equilibrium \cite{WenBook},
but also away from equilibrium (model in Sec.~\ref{Sec. XY with two-body loss}).

In all cases, an effective temperature emerges as the result of dissipation, and the universal behavior including the dynamics near the steady state is generically described by a thermodynamic universality class. The emergent thermal character of driven-dissipative systems may be expected as the quantum coherence is lost to dissipation. However, the phase diagram and the nature of phase transitions, and the precise equivalence with a particular thermal model is often nontrivial, and requires a rather careful treatment based on the Keldysh formalism. This paper offers such a systematic study of four models in great detail and, therefore, illuminates pathways for the beyond-mean-field study of a wide range of other driven-dissipative systems.

We conclude by mentioning other examples of driven-dissipative systems that should be amenable to a similar treatment.
Notable models, also of experimental relevance, are systems with bosons, or photons, coupled to (pseudo-)spins on a lattice. While the two species of fields make the field-theory treatment more complicated, the photonic part is usually quadratic and can be integrated out at the level of the Keldysh action. The resultant effective model is also local due to the dissipative gap of the typically lossy photons.
Examples of experimentally accessible systems of this kind include superconducting circuits \cite{Houck12}, spin-boson networks \cite{Houck14}, strongly interacting Rydberg polaritons \cite{Pritchard10,Dudin12,Peyronel12,Firstenberg13,gunter13}, and internal states of ions coupled to their motion \cite{sorensen00,junemann13}.
We also remark that models closely related to that of Sec.\ \ref{Sec. XY with coherent drive} arise when atoms are coupled to the electromagnetic vacuum, and the latter is eliminated in the Born-Markov approximation \cite{gross82}, a setup that can also be accessed in reduced dimensions \cite{chang12,shahmoon13}.
However, in certain spin-boson-coupled systems, the Born-Markov approximation may not apply, but dissipation caused by external baths may act directly on the bosons and/or the spins.
In this case too, the Keldysh formalism should apply.
It would also be interesting to explore the applicability of the Keldysh formalism  to situations involving dark states \cite{Diehl08,verstraete09} and situations involving transport, as particles or photons continuously enter the system at a boundary \cite{Pritchard10,Peyronel12,petrosyan11,gorshkov13b}.

\begin{acknowledgments}
We thank M.\ Kardar, S. Diehl, M.\ Foss-Feig, A.\ Hu, R. Wilson, Z.-X.\ Gong, M. Hafezi, A. Safavi-Naini, K. Mahmud, and J.\ Young
for discussions. MFM is indebted to M.\ Foss-Feig for many useful discussions. This work was supported by the NSF PIF, AFOSR, ARO, ARL, NSF PFC at the JQI, and AFOSR MURI.
\end{acknowledgments}

\appendix

 \section{Mean field for Model III  A \label{appA}}
 A uniform ansatz for the Model in Sec.~\ref{Sec. TFIM} yields the MF equations
 \begin{align}
   \dot X &= -\hat \Delta\, Y-\frac{\Gamma}{2} X, \nonumber \\
  \dot Y &= \hat \Delta \, X+ \hat J \, X Z-\frac{\Gamma}{2} Y, \\
  \dot Z &= - \hat J\, X Y-\Gamma \,(1+Z), \nonumber
 \end{align}
 where $X=\langle \sx \rangle$, etc. We have defined $\hat \Delta\equiv 2 \Delta$, $\hat J\equiv 2 \z J$, and $\z\equiv2d$ as the coordination number. The MF predicts a continuous transition at
 \begin{equation}
   \Gamma^2 - 4 \hat J \hat \Delta +4 \hat \Delta^2=0,
 \end{equation}
consistent with $r=0$ in the same subsection.

\section{Mean field for Model III B \label{appB}}
A uniform ansatz for the Model in Sec.~\ref{Sec. XY with coherent drive} yields the MF equations
\begin{align}
  \dot X &= -\hat \Delta Y+\hat J \, Y Z-\frac{\Gamma}{2} X, \nonumber \\
  \dot Y &= \hat \Delta X-\hat \Omega \,Z- \hat J \, X Z-\frac{\Gamma}{2} Y, \\
  \dot Z &= \hat \Omega \,Y -\Gamma \,(1+Z). \nonumber
\end{align}
We have defined $\hat \Delta=2\Delta$, $\hat \Omega=2\Omega$, and $\hat J=\z J$.
Casting in terms of $n=(1+Z)/2$, we find for the steady state
\begin{equation}
   \left[4 (\hat \Delta +\hat J-2 \hat J \,n)^2+2 \hat\Omega ^2+\Gamma^2\right] n= \Omega ^2\,,
\end{equation}
which is similar to the mean-field equation in the continuum, Eq.~(\ref{Eq. eqn for psi0}).
The MF equation above exhibits a continuous transition from a stable uniform phase to a bistable region with two stable uniform phases.
\\

\section{Mean field for Model III C \label{appC}}

The MF for the model in Sec.~\ref{Sec. Anisotropic XY} is derived in Ref.~\cite{Lee13}. A two-site ansatz on the checkerboard sublattices $A$ and $B$ yields the mean field equations
\begin{align}
  \dot X_A &= -2 J Z_A Y_B -\frac{\Gamma}{2} X_A, \nonumber \\
  \dot Y_A &= -2 J Z_A X_B -\frac{\Gamma}{2} Y_A, \\
  \dot Z_A &= 2 J (Y_A X_B-X_A Y_B) -{\Gamma} (1+Z_A), \nonumber
\end{align}
and a similar set of equations for $A\leftrightarrow B$. The MF predicts a continuous phase transition from a paramagnetic state with $Z=-1$ to a staggered XY phase where spins on the two sublattices are at angles $\theta$ and $-\theta$ with respect to the $x=y$ line on the Bloch sphere, see Ref.~\cite{Lee13} and the figure therein. The phase transition occurs at
\begin{equation}
   J=\frac{\Gamma}{4}\,,
\end{equation}
consistent with setting $r=0$ in Sec.~\ref{Sec.   Anisotropic XY}.

 \section{Mean field for Model III D \label{appD}}
 A uniform ansatz for the Model in Sec.~\ref{Sec. XY with two-body loss} yields the MF equations [$n=(1+\langle \sz\rangle)/2$]
 \begin{align}
   \dot X &= -\left[\hat J\,(2n -1)+\hat \Delta\right] Y-\left(\frac{\Gamma+\Gamma_p}{2}+\kappa \, n\right) X, \nonumber \\
  \dot Y &= \left[\hat J\,(2n -1)+\hat \Delta\right] X-\left(\frac{\Gamma+\Gamma_p}{2}+\kappa \, n\right) Y,\\
  \dot n &= - \Gamma n+\Gamma_p \, (1-n) -\kappa \, n^2, \nonumber
 \end{align}
 where $\hat J=2\z J$ and $\hat \Delta=2\Delta$.
 In the steady state, $n$ varies between $0$ and $1$ depending on decay rates; however, the MF always gives $X=Y=0$.

\end{document}